\documentclass[aps,twocolumn,prb,floatfix,superscriptaddress,longbibliography]{revtex4-1}
\usepackage{graphicx}
\usepackage[latin9]{inputenc}
\usepackage{amsmath,amsfonts,amssymb,bm}

\usepackage{lmodern}
\usepackage{xcolor}
\usepackage[normalem]{ulem}
\usepackage{float}
\usepackage{braket}
\usepackage{dsfont}

\usepackage[colorlinks,
            linkcolor=blue,    
            citecolor=blue,  
            urlcolor=blue,
 	        bookmarks=true,        
	        bookmarksopen=true,    
	        bookmarksnumbered=true,
]{hyperref}

\usepackage{ulem}
\global\long\def\ket#1{\left| #1\right\rangle }%

\global\long\def\bra#1{\left\langle #1 \right|}%

\global\long\def\tr{\text{tr}}%

\global\long\def\pd{\partial}%

\global\long\def\abs#1{\left|#1\right|}%

\global\long\def\bs#1{\boldsymbol{#1}}%

\global\long\def\t#1{\text{#1}}%

\global\long\def\bs#1{\boldsymbol{#1}}%

\usepackage{bm}
\usepackage{braket}

\makeatother

\begin{document}

\title{Symmetry-induced even/odd parity in charge and heat pumping}

\author{Miguel A. N. Ara\'ujo}
\affiliation{CeFEMA, Instituto Superior T\'{e}cnico, Universidade de Lisboa, Av. Rovisco
Pais, 1049-001 Lisboa, Portugal}
\affiliation{Departamento de F\'{\i}sica, Universidade de  \'Evora, P-7000-671, \'Evora,
Portugal}
\affiliation{Beijing Computational Science Research Center, Beijing 100084, China}
\author{Pedro Ninhos}
\affiliation{CeFEMA, Instituto Superior T\'ecnico, Universidade de Lisboa, Av. Rovisco
Pais, 1049-001 Lisboa, Portugal}
\author{Pedro Ribeiro}
\affiliation{CeFEMA, Instituto Superior T\'ecnico, Universidade de Lisboa, Av. Rovisco
Pais, 1049-001 Lisboa, Portugal}
\affiliation{Beijing Computational Science Research Center, Beijing 100084, China}

\begin{abstract}
We study the effect of discrete symmetries on charge and heat pumping through 
non-interacting
Floquet systems when spatial symmetry (SS) is broken. 
Particle-hole symmetry (PHS) implies that the pumping
of charge (heat) is an odd (even) function of the chemical potential.
If PHS is broken, the product of PHS and SS produces
even (odd) charge (heat) pumping. Simultaneous breaking of PHS and
SS can be due to the coupling to the leads, even if the latter
is spatially symmetric. This provides a very simple criterion for
reversing (or maintaining) the direction of the flow. We illustrate these
results by considering two variants of the Su-Schrieffer-Heeger model
under a time-periodic perturbation. 
\end{abstract}
\maketitle

\section{Introduction}

Quantum pumping consists of coherently transporting heat or charge between
macroscopic reservoirs using a periodic drive - the pump.  
By exchanging energy with the drive, carriers can be transported across reservoirs even
in the absence of any temperature or voltage bias. 

The pumping of charge was first understood in the so-called adiabatic limit
\citep{Brouwer,Buttiker1994}, where the ratio 
between the driving period, $T=2\pi/\Omega$, and the time taken by
carriers to traverse the sample, $\tau$, is large, i.e. $2\pi/\Omega\tau\gg1$.
As $\tau$ is proportional to the linear sample size, $L$, the adiabatic
approximation eventually breaks down and one has to resort to a fully
time dependent modeling using scattering Floquet theory \citep{buttiker,Arrachea2006}
or a non-equilibrium Green's functions approach \citep{Arrachea2006,Wu2008,Stefanucci2008,Wu2010,Cabra2020,Gaury2014}
. Since then, quantum charge pumps have found
applications in quantum metrology \citep{Giblin2012,Connolly2013},
single-photon or electron emitters \citep{Mueller2010,Bocquillon2013}
and quantum information processing \citep{Trauzettel2007}.

It has long been noted that pumping is only allowed when spatial symmetry is
broken  \citep{buttiker,hanggi}.
However,  a systematic understanding of how symmetry affects the pumping of heat and charge is still incomplete.
In this article, we take a step further in this direction
 by showing that some symmetry properties leave a signature on the oven/odd dependence of the charge and heat pumped, within a cycle, on the reservoirs chemical potential. 
Specifically, we show that if inversion symmetry along the direction of the current
 is broken and particle-hole symmetry (PHS) holds, or if only the product
of the two holds, charge and heat pumping  then depend on the chemical potential
in qualitatively different ways: the pumped charge or heat in one cycle can either be an even or odd function of the chemical potential. 
We extend these findings to compositions of the above symmetries with unitary symmetries.  
This result provides a very simple criterion for reversing (or maintaining) the direction of the flow. 

We illustrate our findings in two variants of the 
 periodically driven Su-Schrieffer-Heeger
(SSH) chain \cite{SSHum,SSHdois} coupled to two wide-band reservoirs  \citep{hanggi}.  
Symmetry breaking can either be induced by the coupling to the leads or
by explicitly breaking spatial symmetry, for instance, through a spatially non-uniform drive of the 1D conductor. 
 As an example of a composite symmetry, we
introduce a special form of PHS, acting in space-time domain, which  affects differently the two model variants.

The paper is organized as follows. We define the setup in Section \ref{SecPumping} and recall
the charge and energy currents within the Floquet Green's function
formalism. 
In Section \ref{SecSymmetries}  
we discuss the relevant symmetries, how they apply to the Floquet scattering matrix
and Green's function, and derive their implications for the
transmission probabilities.
In Section \ref{SecTransport} we derive their implications for
the average pumped heat and charge per cycle.
We present examples in Section \ref{SecExamples}.
We summarize and discuss the implications of our findings in Section \ref{SecSummary}.

\section{Charge and energy pumping} 
\label{SecPumping}

We consider a typical
transport setup consisting of two macroscopic metallic leads connected
by a mesoscopic system, $\t S$, to which the driving is applied.
The total Hamiltonian is given by 
\begin{align}
H\left(t\right) & =H_{\t S}\left(t\right)+\sum_{l=\t{R,L}}\left(H_{l}+H_{l-\t S}\right),\label{eq:Hamiltonian}
\end{align}
where the Hamiltonian of the system, 
$H_{\t S}\left(t\right)=\sum_{\alpha\beta}\hat c_{\alpha}^{\dagger}\mathbb{H}_{\t S;\alpha\beta}\left(t\right)\hat c_{\beta}$,
is assumed to be quadratic, with $\hat c_{\alpha}^{\dagger}$ and $\hat c_{\alpha}$
the fermionic creation and annihilation operators, with a periodic
single-particle Hamiltonian, 
$\mathbb{H}_{\t S}\left(t\right)=\mathbb{H}_{\t S}\left(t+T\right)$. 
The Hamiltonians, $H_{l}$,
for the right and left $(l=\t R,L)$  leads  are time independent and non-interacting, 
and the same applies to the system-lead
coupling term, $H_{l-\t S}$.
Under these conditions, the retarded Green's function of the system
verifies Dyson's equation, $\left[i\pd_{t}-\mathbb{H}_{\t S}\left(t\right)\right]G^{R}=\sum_{l}\Sigma_{l}^{R}.G^{R}$,
 where $\Sigma_{l}^{R}\left(t,t'\right)=\int\frac{d\omega}{2\pi}e^{-i\omega\left(t-t'\right)}\Sigma_{l}^{R}\left(\omega\right)$
is the time-translational invariant retarded self-energy induced by
lead $l$. Under periodic driving, it is convenient to define the
Floquet Green's function \cite{hanggi}:
\begin{align}
\mathcal{G}_{\left(m\right)}\left(\epsilon\right) & =\frac{1}{T}\int_{0}^T dt\int_{-\infty}^{+\infty}d\tau e^{i\left(m\hbar\Omega t-\epsilon\tau\right)}G^{R}\left(t,t+\tau\right).\label{eq:Floquet_GF}
\end{align}
Assuming there are no bound-states, at large times after the periodic
drive has been turned on, a recurrent state is attained and observables
become periodic with driving period \cite{Stefanucci2008}. Here,
we are concerned with the average charge $J_l^{c}$, and energy, $J_l^{e}$,
currents leaving lead $l$ over one driving cycle, defined as $J_l^{c/e}=\lim_{\tau\to\infty}\int_{\tau}^{\tau+T}\frac{dt}{T}J_l^{c/e}\left(t\right)$.
In terms of the Floquet Green's function, average currents are given
by\citep{hanggi,Stefanucci2008} 
\begin{align}
J_{l}^{c} & =-\frac{|e|}{h}\sum_{m\in\mathbb{Z}}\int d\epsilon\left\{ T_{l\bar{l}}^{\left(m\right)}\left(\epsilon\right)f_{\bar{l}}\left(\epsilon\right)-T_{\bar{l}l}^{\left(m\right)}\left(\epsilon\right)f_{l}\left(\epsilon\right)\right\} ,\label{eq:charge_J}\\
J_{l}^{e} & =\frac{1}{h}\sum_{m\in\mathbb{Z}}\int d\epsilon\left\{ \left(m\hbar\Omega+\epsilon\right)T_{l\bar{l}}^{\left(m\right)}\left(\epsilon\right)f_{\bar{l}}\left(\epsilon\right)\right.\nonumber \\
 & \hspace{1cm}\left.-\epsilon T_{\bar{l}l}^{\left(m\right)}\left(\epsilon\right)f_{l}\left(\epsilon\right)+m\hbar\Omega T_{ll}^{\left(m\right)}\left(\epsilon\right)f_{l}\left(\epsilon\right)\right\} \,,\label{eq:energy_J}
\end{align}
where,
$T_{ll'}^{\left(m\right)}\left(\epsilon\right)=\tr\left[\mathcal{G}_{\left(m\right)}\left(\epsilon\right)\Gamma_{l'}\left(\epsilon\right)\mathcal{G}_{\left(m\right)}^{\dagger}\left(\epsilon\right)\Gamma_{l}\left(\epsilon+m\hbar\Omega\right)\right]$
is the transmission probability for a fermion leaving $l'$ with energy $\epsilon$ 
and arriving at $l$ after absorbing $m$  energy quanta (photons) from the driving field.
$\Gamma_{l}\left(\epsilon\right)=i\left[\Sigma_{l}^{R}\left(\epsilon\right)-\Sigma_{l}^{A}\left(\epsilon\right)\right]$
is the hybridization matrix of lead $l$, and we introduced the notation $\bar{\t R}=\t L$ and $\bar{\t L}=\t R$. 
$f_l\left(\epsilon\right)$ denotes the Fermi-Dirac distribution function at lead $l$ at temperature $1/\beta$.
The first term of Eq.(\ref{eq:energy_J}) describes the energy absorbed by $l$ when an electron leaves
$\bar{l}$ with energy $\epsilon$ and absorbs $m$ photons; the
second term is the energy lost by $l$ when a electron with energy
$\epsilon$ is transmitted to $\bar{l}$; and the last term is the
energy gained when an electron is reflected back to $l$ having absorbed
$m$ photons. 

In the following, we set the reservoirs to the same chemical potential, $\mu_l=\mu$,
and consider the total charge transferred in one cycle between the
leads, $Q=\frac{2\pi}{\Omega}J_{L}^{c}$. Charge conservation ensures
that $J_{\t L}^{c}=-J_{\t R}^{c}$. 
We also consider the total energy
generated in one cycle, $\mathcal{E}_{t}=\frac{2\pi}{\Omega}\left(J_{\t L}^{e}+J_{\t R}^{e}\right)$,
and the energy pumped between leads, $\mathcal{E}_{\Delta}=\frac{2\pi}{\Omega}\left(J_{\t L}^{e}-J_{\t R}^{e}\right),$
as functions of the leads' chemical potential. For convenience,
we study the derivatives of these quantities with respect to $\mu$, 
\begin{eqnarray}
Q'\left(\mu\right)  &=&\int \frac{\beta d\epsilon}{4\cosh^{2}\left[\frac{\beta(\epsilon-\mu)}{2}\right]}Q^{(0)'}\left(\epsilon\right), \label{Qbeta}\\
\mathcal{E}'_{t/\Delta}\left(\mu\right) &=& \int \frac{\beta d\epsilon}{4\cosh^2\left[\frac{\beta(\epsilon-\mu)}{2}\right]}
\mathcal{E}^{(0)'}_{t/\Delta}\left(\epsilon\right),
\end{eqnarray}
where the corresponding zero-temperature expressions read
\begin{align}
Q^{(0)'}\left(\epsilon\right)  &=-\frac{|e|}{\Omega\hbar} 
\sum_{m\in\mathbb{Z}}\left[T_{\t{LR}}^{\left(m\right)}\left(\epsilon\right)-T_{\t{RL}}^{\left(m\right)}\left(\epsilon\right)\right],\label{eq:Q'}\\
\mathcal{E}^{(0)'}_{t/\Delta}\left(\epsilon\right) & = \frac{1}{h}\sum_{m\in\mathbb{Z}} \bigg\{ \sum_l \left[T_{\t{L}l}^{\left(m\right)}\left(\epsilon\right)\pm T_{\t{R}l}^{\left(m\right)}\left(\epsilon\right)\right]m \nonumber \\
 & \hspace{0.1cm} +(1\mp1 )\frac{\epsilon}{\hbar\Omega}\left[T_{\t{LR}}^{\left(m\right)}\left(\epsilon\right)\pm T_{\t{RL}}^{\left(m\right)}\left(\epsilon\right)\right] \bigg\} . \label{eq:QE'}
\end{align}
 
The heat current follows from (\ref{eq:energy_J}) as the transport of $\epsilon-\mu$ instead of $\epsilon$, 
\begin{eqnarray}
J_{l}^{h} & = J_{l}^{e} - \frac{\mu}{\abs{e}} J_{l}^{c}.\label{eq:heat_J}
\end{eqnarray}
So, the heat transported per cycle  is given by
$\mathcal{Q}_{t}=  \mathcal{E}_{t} $ and $\mathcal{Q}_{\Delta}= \mathcal{E}_{\Delta}  - \frac{\mu}{\abs{e}} 2 Q$.

Expressions (\ref{eq:charge_J}) and (\ref{eq:energy_J})
have also been derived using the Floquet scattering matrix approach~\cite{buttiker}.  
We shall next study how the system's symmetries affect the transmission probabilities: first, by
considering the scattering matrix; later on,  within the Green's function formalism. 
The latter allows us to integrate out the leads and obtain a non-hermitian Hamiltonian for the system, 
for which a recently proposed set of symmetries~\cite{SatoSYM} applies.

\section{Symmetries} \label{SecSymmetries}

We employ the Altland-Zirnbauer (AZ) classification 
of hermitian operators according to discrete symmetries~\citep{ludwig2008,Hasan2010,Bernevig2013}
and consider the transformation, $\mathbb{H}\left(t\right)\to\mathbb{H}_{X}\left(t\right)$,
of the full Hamiltonian (of $\t S$ and leads), under the  symmetry
transformation, $X$.  The equalitiy $\mathbb{H}_{X}=\mathbb{H}$ holds whenever the symmetry $X$ is present.

Time-reversal and particle-hole transformations read
$\mathbb{H}_{T}\left(t\right)=\mathbb{U}_{T}^{\dagger}\mathbb{H}^*\left(-t\right)\mathbb{U}_{T}$
and $\mathbb{H}_{C}\left(t\right)=-\mathbb{U}_{C}^\top\mathbb{H}^*\left(t\right)\mathbb{U}_{C}^{*}$,
respectively,
where $\mathbb{U}_{X}$ are suitable unitary matrices.
We also consider $x$-axis inversion, 
$\bra x\mathbb{H}_{P}\left(t\right)\ket{x'}=\mathbb{U}_{P}^{\dagger}\bra{-x}\mathbb{H}\left(t\right)\ket{-x'}\mathbb{U}_{P}$,
where  $x$ is the coordinate along the propagation direction
of the current.
As in one dimension $x$-axis inversion coincides with parity symmetry (PS), we denote this transformation as $P$. 
However, the arguments below are valid for any symmetry transformation that inverts the $x$-axis.

We also consider the half-period time translation transformation,
$\mathbb{H}_{\Pi}\left(t\right)=\mathbb{U}_{\Pi}^{\dagger}\mathbb{H}\left(t+T/2\right)\mathbb{U}_{\Pi}$.
Finally, for a generic local unitary transformation, $K$, acting only
on the unit cell,  $\mathbb{H}_{K}\left(t\right)=\mathbb{U}_{K}^{\dagger}\mathbb{H}\left(t\right)\mathbb{U}_{K}$.
A particular case is that of the local operator, $\mathbb{U}$, which reads, in real space,
 $\mathbb{U}|x\rangle=\sigma_3(-1)^x|x\rangle$,
and will be used in Section~\ref{SecExamples}.

\subsection{Symmetries of the  Floquet scattering matrix}

The implications of the above symmetries for  
the transport properties can be obtained most simply 
within the Floquet scattering matrix approach~\cite{buttiker,reichl}. 
We next discuss the symmetry properties of the Floquet scattering matrix
in general terms, where the asymptoptic form of the wave function far from a scatterer assumes a plane-wave form.

We view the Floquet function 
\begin{equation}
\psi(x,t)=\sum_n\phi_n(x) e^{-\frac{i}{\hbar}Et}e^{-in\Omega t}\,,
\end{equation}
with quasi-energy $E$,   
as a superposition of states with energies $E+n\hbar\Omega$. 
For a  scattering state, the spatial part of the Floquet functions,  $\phi_n(x)$, 
takes the form of plane waves far from the scatterer:
 \begin{eqnarray}
\phi_n(x\rightarrow -\infty) &=& A_n e^{ikx} + C_n  e^{-ikx}\\
\phi_n(x\rightarrow +\infty) &=& D_n e^{ikx} + B_n  e^{-ikx}
\,.
\end{eqnarray}
The Floquet scattering matrix, $S(E+n'\hbar\Omega, E+n\hbar\Omega)$, 
relates the Fourier amplitudes of the incoming waves 
with the outgoing ones: 
 \begin{eqnarray}
\left[\begin{array}{c}  C_{n'} \\  D_{n'}   \end{array} \right] = S(E+n'\hbar\Omega, E+n\hbar\Omega)
\left[\begin{array}{c}  A_n\\  B_n  \end{array} \right]
\,. \label{Sdefin}
\end{eqnarray}
We may think of the column vectors as having all entries  $n\in \mathbb{Z}$. Then, the $S$ matrix
has four blocks: 
 \begin{eqnarray}
S=\left[\begin{array}{cc} 
S_{LL}(n',n) & S_{LR}(n',n) \\ S_{RL}(n',n) & S_{RR}(n',n) \end{array} \right]\,,
\label{Sblocks}
\end{eqnarray}
and we rewrite (\ref{Sdefin}) as
 \begin{eqnarray}
\left[\begin{array}{c}  C^*\\  D^*  \end{array} \right] = S\
\left[\begin{array}{c}  A^* \\  B^*   \end{array} \right] 
\,.
\end{eqnarray}
Probability conservation implies $SS^\dagger =1$.  The relation between the $S$ matrix and 
the above transmission probabilities is
\begin{equation}
T_{ll'}^{\left(m\right)}\left(\epsilon\right) =|S_{ll'}(\epsilon+m\hbar\Omega, \epsilon)|^2\,.
\label{TSrelation}
\end{equation}
We present in Appendix \ref{ApFloS} the derivation of the following symmetry properties of the scattering matrix:  
\begin{enumerate}
\item  TRS implies that 
\begin{equation}
S^\top=S\,.
\label{TRSmatrix}
\end{equation}
 \item PHS implies that
   \begin{eqnarray}
 S^*(E+n'\hbar\Omega,E) &=&  S(-E-n'\hbar\Omega,-E) 
 \,.
 \label{PHSmatrix}
 \end{eqnarray}
\item Parity symmetry implies that 
 \begin{eqnarray}
 S_{LL}(n',n) &=& S_{RR}(n',n)\,, \label{SLLsymm}\\
 S_{LR}(n',n) &=& S_{RL}(n',n) \label{SLRsymm}
 \,.
\end{eqnarray}

 \item Symmetry under $\mathbb{U} \Pi C$ operator, where
 $\mathbb{U}|x\rangle=\sigma_3(-1)^x|x\rangle$, in real space, implies that  
 \begin{eqnarray}
\begin{split}
& S^{(\nu)^*}(E+n'\hbar\Omega,E)=\\ 
&S^{(-\nu)}(-E-n'\hbar\Omega,-E)(-1)^{n'} \,.\\
\end{split}
 \end{eqnarray}
 
\end{enumerate}

From the above symmetry properties of the scattering matrix and Eq.(\ref{TSrelation}), 
the symmetry properties of the transmission probabilities can be obtained.

\subsection{Green's functions and their symmetries}

We now analyse the symmetry properties
of the Green's functions that follow from those of the Hamiltonian. 
Since the total system evolves unitarily, the transformed Green's
function is given by $G_{X}^{R}\left(t,t'\right)=-i\Theta\left(t-t'\right)\mathcal{T}e^{-i\int_{t'}^{t}d\tau\,\mathbb{H}_{X}\left(\tau\right)}$. Using the transformations $\mathbb{H}_X$, 
the transformed Green's functions read (see Appendix \ref{ApA})
\begin{align}
G_{T}^{R}\left(t,t'\right) & =\mathbb{U}_{T}^{\dagger}\left[G^{R}\left(-t',-t\right)\right]^\top\mathbb{U}_{T},\label{eq:t_rules_1}\\
G_{C}^{R}\left(t,t'\right) & =-\mathbb{U}_{C}^\top\left[G^{R}\left(t,t'\right)\right]^{*}\mathbb{U}_{C}^{*},\\
\bra xG_{P}^{R}\left(t,t'\right)\ket{x'} & =\bra{-x}\mathbb{U}_{P}^{\dagger}G^{R}\left(t,t'\right)\mathbb{U}_{P}\ket{-x'}\,.\label{eq:t_rules_2}
\end{align}
 
The equalitiy $G_{X}^{R}\left(t,t'\right)=G^{R}\left(t,t'\right)$ holds whenever the symmetry $X$ is present.

If the unitary matrices, $\mathbb{U}_{X}$, do not mix degrees of
freedom of the system with those of the leads, we may define 
Green's functions restricted to  the degrees of freedom solely within the system as
 $G_{\t S}^{R}=\sum_{\alpha,\beta\in\t S}\ket{\alpha}\bra{\alpha}G^{R}\ket{\beta}\bra{\beta}$.
Then, $G_{\t S}^{R}$ follows the same transformation rules and obeys
\begin{multline}
\left[i\pd_{t}-\mathbb{H}_{X,\t S}\left(t\right)\right]G_{X,\t S}^{R}\left(t,t'\right)=\\
\int d\tau\sum_{l=\t{R,L}}\Sigma_{X,l}^{R}\left(t,\tau\right)G_{X,\t S}^{R}\left(\tau,t'\right),
\end{multline}
where $\Sigma_{X,l}^{R}$ is the transformed self-energy of lead $l$
that transforms as $G_{X,\t S}^{R}$. 

We note that when $\Sigma_{l}^{R}\left(t,t'\right)=\delta\left(t-t'\right)\Sigma_{l}^{R}\left(t\right)$,
the Green's function can be written as $G_{\t S}^{R}\left(t,t'\right)=-i\Theta\left(t-t'\right)\mathcal{T}e^{-i\int_{t'}^{t}d\tau\,Z\left(\tau\right)}$,
where $Z=\mathbb{H}_{\t S}\left(t\right)+\sum_{l}\Sigma_{l}^{R}\left(t\right)$ can be identified with an effective non-hermitian
Hamiltonian, which transforms as
 $Z_{T}\left(t\right)=\mathbb{U}_{T}^{\dagger}Z^\top\left(-t\right)\mathbb{U}_{T}$
and $Z_{C}\left(t\right)=-\mathbb{U}_{C}^\top Z^{*}\left(t\right)\mathbb{U}_{C}^{*}$.
The symmetries associated with these transformations also arise in Markovian environments~\cite{Lieu2020}, and
 were recently proposed under the nomenclature TRS$^\dagger$ and PHS$^\dagger$, respectively,
  in Ref.~\cite{SatoSYM}. 

In the following, we drop the label
$\t S$ and refer to $G^{R}$ as the Green's function of the system. 

Using Eq.(\ref{eq:Floquet_GF}) and the transformation rules (\ref{eq:t_rules_1}-\ref{eq:t_rules_2})
it is straightforward to show that
\begin{align}
\mathcal{G}_{T;\left(m\right)}\left(\epsilon\right) & =\mathbb{U}_{T}^{\dagger}\left[\mathcal{G}_{\left(-m\right)}\left(\epsilon+m\hbar\Omega\right)\right]^\top\mathbb{U}_{T},\label{t}\\
\mathcal{G}_{C;\left(m\right)}\left(\epsilon\right) & =-\mathbb{U}_{C}^\top\left[\mathcal{G}_{\left(-m\right)}\left(-\epsilon\right)\right]^{*}\mathbb{U}_{C}^{*},\\
\bra x\mathcal{G}_{P;\left(m\right)}\left(\epsilon\right)\ket{x'} & =\mathbb{U}_{P}^{\dagger}\bra{-x}\mathcal{G}_{\left(m\right)}\left(\epsilon\right)\ket{-x'}\mathbb{U}_{P}\,.\label{p}
\end{align}
In turn,  Eqs. (\ref{t}-\ref{p}) can be used to deduce the transformation properties of the transmission probabilities.
The details of the derivation are given in Appendices \ref{ApA} and \ref{ApB}. 
They read
\begin{align}
T_{T;ll'}^{\left(m\right)}\left(\epsilon\right)= & T_{l'l}^{\left(-m\right)}\left(\epsilon+m\hbar\Omega\right)\label{eq:T_T}\,,\\
T_{C;ll'}^{\left(m\right)}\left(\epsilon\right)=  T_{ll'}^{\left(-m\right)}\left(-\epsilon\right)
&\,,\hspace{0.5cm}
T_{P;ll'}^{\left(m\right)}\left(\epsilon\right)= T_{\bar l \bar l'}^{\left(m\right)}\left(\epsilon\right)\label{eq:T_P}\,.
\end{align}
Eq.(\ref{eq:T_T}) can be also obtained from 
Eqs(\ref{TSrelation}) and  (\ref{TRSmatrix}). 
And Eq.(\ref{eq:T_P}) can be obtained from Eqs(\ref{TSrelation}), (\ref{PHSmatrix}),
(\ref{SLLsymm})-(\ref{SLRsymm}).

\section{Implications for transport}
\label{SecTransport}

\subsection{Parity and particle-hole symmetry}

We now derive the implications
of the above symmetries for transport.
It has long been known that charge, energy, or heat
pumping requires inversion symmetry breaking~\citep{buttiker,hanggi}. 
If the $x$-axis inversion leaves the full Hamiltonian, $\mathbb{H}$, invariant, then 
the Green's function remains
invariant, $\mathcal{G}_{P;\left(m\right)}\left(\epsilon\right)=\mathcal{G}_{\left(m\right)}\left(\epsilon\right)$,
whereas the hybridizations are interchanged,
$\Gamma_{P;l}\left(\epsilon\right)=\Gamma_{\bar{l}}\left(\epsilon\right)$.
In this case, $T_{ll'}^{\left(m\right)}\left(\epsilon\right)=T_{\bar{l}\bar{l'}}^{\left(m\right)}\left(\epsilon\right)$,
which can also be derived from Eqs.(\ref{TSrelation}), (\ref{SLLsymm})-(\ref{SLRsymm}).
This symmetry property implies that
$Q'\left(\mu\right)=\mathcal{Q}'_{\Delta}\left(\mu\right)=0$, 
so no transport of charge or heat occurs.

However, if the system plus reservoirs are invariant under PHS, then
$\mathcal{G}_{C;\left(m\right)}\left(\epsilon\right)=\mathcal{G}_{\left(m\right)}\left(\epsilon\right)$
and $\Gamma_{C;l}\left(\epsilon\right)=\Gamma_{l}\left(\epsilon\right)$.
This implies (see Appendix \ref{ApB})
\begin{eqnarray}
T_{ll'}^{\left(m\right)}\left(\epsilon\right)=T_{C;ll'}^{\left(m\right)}\left(\epsilon\right)=T_{ll'}^{\left(-m\right)}\left(-\epsilon\right)\,,
\end{eqnarray}
which also follows from Eqs.(\ref{TSrelation}) and (\ref{PHSmatrix}). 
From Eqs.(\ref{Qbeta})-(\ref{eq:QE'}), we now obtain
\begin{eqnarray}
 Q'\left(\mu\right)=Q'\left(-\mu\right), \ \ 
\mathcal{Q}'_{t/\Delta}\left(\mu\right)=-\mathcal{Q}'_{t/\Delta}\left(-\mu\right),
\end{eqnarray}
Then, $Q\left(\mu\right)$ is an odd function and $\mathcal{Q}_{t/\Delta}\left(\mu\right)$ are even. 
This is because $Q'(\mu_l)$ even implies $Q(\mu_l)$ odd plus a constant. That this constant
is zero  can be seen by considering either the limit $\mu_l\rightarrow -\infty$, where no available
particles exist, or the opposite limit $\mu_l\rightarrow \infty$, where the fermionic states are all occupied
and, therefore, Pauli blocked.

\subsection{Composition of symmetries}

It may happen that both $x$-axis inversion and PHS are broken while their product,  $\mathcal{PC}$, still holds as a symmetry.
In that case,
 $T_{ll'}^{\left(m\right)}\left(\epsilon\right)=T_{PC;ll'}^{\left(m\right)}\left(\epsilon\right)=T_{\bar{l}\bar{l'}}^{\left(-m\right)}\left(-\epsilon\right)$,
and from Eqs.(\ref{Qbeta})-(\ref{eq:QE'}),
 \begin{eqnarray}
 Q'\left(\mu\right)&=&-Q'\left(-\mu\right),\label{Qlodd}\\ 
 \mathcal{Q}'_{t}\left(-\mu\right)=-\mathcal{Q}'_{t}\left(\mu\right), &&
 \mathcal{Q}'_{\Delta}\left(\mu\right)=\mathcal{Q}'_{\Delta}\left(-\mu\right)\label{Qdeltaeven}.
 \end{eqnarray}
In this case 
 the pumped charge (heat) is an even (odd) function of $\mu$ and the total heat absorbed is even.

Composition with other unitary symmetry leads to the same even/odd pumping relations versus $\mu$.
Consider, for instance, the half-period time translation, ${\Pi}$,
the generic unitary symmetry implemented by an unitary operator, $\mathbb{U}$, that acts locally on the unit cells.
Under the composition 
 $\Pi C$, we have $\mathcal{G}_{\Pi C;\left(m\right)}\left(\epsilon\right)=\left(-1\right)^{m}\mathbb{U}_{\Pi}^{\dagger}\mathcal{G}_{C;\left(m\right)}\left(\epsilon\right)\mathbb{U}_{\Pi}$
and $\Gamma_{\Pi C}\left(\epsilon\right)=\mathbb{U}_{\Pi}^{\dagger}\Gamma_{C}\left(\epsilon\right)\mathbb{U}_{\Pi}$,
implying
$T_{\Pi C;ll'}^{\left(m\right)}\left(\epsilon\right)=T_{C;ll'}^{\left(m\right)}\left(\epsilon\right)$.
Therefore, for a system invariant under $\Pi C$, both $Q'\left(\mu\right)$
and $\mathcal{Q}'_{t/\Delta}\left(\mu\right)$ have the same properties
under $\mu\to-\mu$ as a system invariant under $C$. 
In the same way, one can show that invariance under the combination $\Pi PC$, yields the same results
as invariance under $PC$. 

More generally, for $X=\mathbb{U},\Pi,  \mathbb{U}\Pi$, 
invariance under the combination $XP$, $XC$ and $XPC$,  yields the same results as invariance under $P$, $C$ and $PC$,
respectively. 
An account of the symmetries and their effects of the different pumping quantities is given in Table~\ref{default}.

\subsection{The role of  time reversal symmetry}

For the transmission
probability, time reversal symmetry (TRS) implies 
$T_{ll'}^{\left(m\right)}\left(\epsilon\right)=T_{l'l}^{\left(-m\right)}\left(\epsilon+m\hbar\Omega\right)$
[see Eqs.(\ref{TSrelation}), (\ref{TRSmatrix}) and (\ref{eq:T_T}) ].
Although these probabilities are the same, these two processes will happen
at different rates due to the different occupation numbers of energies
$\epsilon$ and $\epsilon+m\hbar\Omega$ in the equilibrium distribution of the leads. Therefore, only for $f\left(\epsilon+m\hbar\Omega\right)=f\left(\epsilon\right)$,
which requires infinite temperature ($f\left(\epsilon\right)=\frac{1}{2}$),
can TRS be used to infer qualitative features of
transport quantities. It is then easy to show that breaking
TRS allows for pumping 
between infinite temperature
leads whereas all currents vanish in the time-symmetric case (see Appendix~\ref{ApC}).

\begin{table}[htp]
\begin{center}
\begin{tabular}{c|c|c|c|c|lc}
\multicolumn{3}{c|}{Symmetry} & \multicolumn{2}{c|}{Pumping}  \\
  PHS  & PS & $\cal PC$  & $Q(\mu)$  & ${\cal Q}_{\Delta/t}(\mu)$   & Model example\\
\hline
\hline
$C$  & $P$ & $PC$ & 0      & 0/even & $ Z_{xy}$ (hom)    \\
$\mathbb{U}\Pi C$ & $P$ & $P\mathbb{U}\Pi C$ & 0  & 0/even & $\left[Z_{zx}\right]_{\nu= 0}$ (hom)   \\
\hline
$C$ & - & -   & odd & even/even & $Z_{xy}$ (inhom) \\
$\mathbb{U}\Pi C$& - & -   & odd  & even/even & $\left[Z_{xy}\right]_{\nu= 0}$ (inhom)\\
$\mathbb{U}\Pi C$& - &  -  & odd & even/even & $\left[ Z_{zx}\right]_{\nu=0}$  (inhom)  \\
\hline
 - & - & $PC$ & even & odd/even & $\left[Z_{zx}\right]_{\nu\neq 0}$ (hom)  \\
\hline
\end{tabular}
\end{center}
\caption{Model symmetries and parity of charge and heat pumping. 
PS denotes $x$-axis inversion, and $\cal PC$ the composition of PHS and PS.
Examples are included where PHS is implemented by a transformation $XC$ with $X=\mathbb{U}\Pi$ 
 [See Eqs.(\ref{OHzx})-(\ref{OSigma})] . }
\label{default}
\end{table}

\section{Examples}
\label{SecExamples}

\begin{figure}[h]
\centering
\begin{minipage}[b]{1\linewidth}\hspace{-3.9cm}
\includegraphics[width=0.4\linewidth]{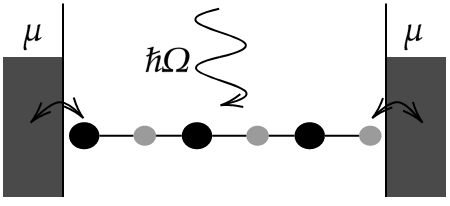}
\end{minipage}%
\begin{minipage}[b]{1\linewidth}\hspace{-12.3cm}\vspace{0.0cm}
\includegraphics[width=0.4\linewidth]{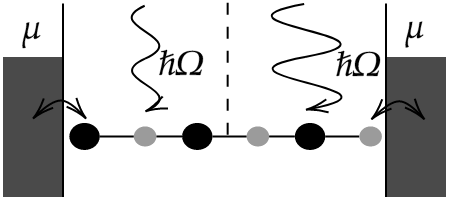}
\end{minipage}
\begin{minipage}[b]{1\linewidth}
\includegraphics[width=1\linewidth]{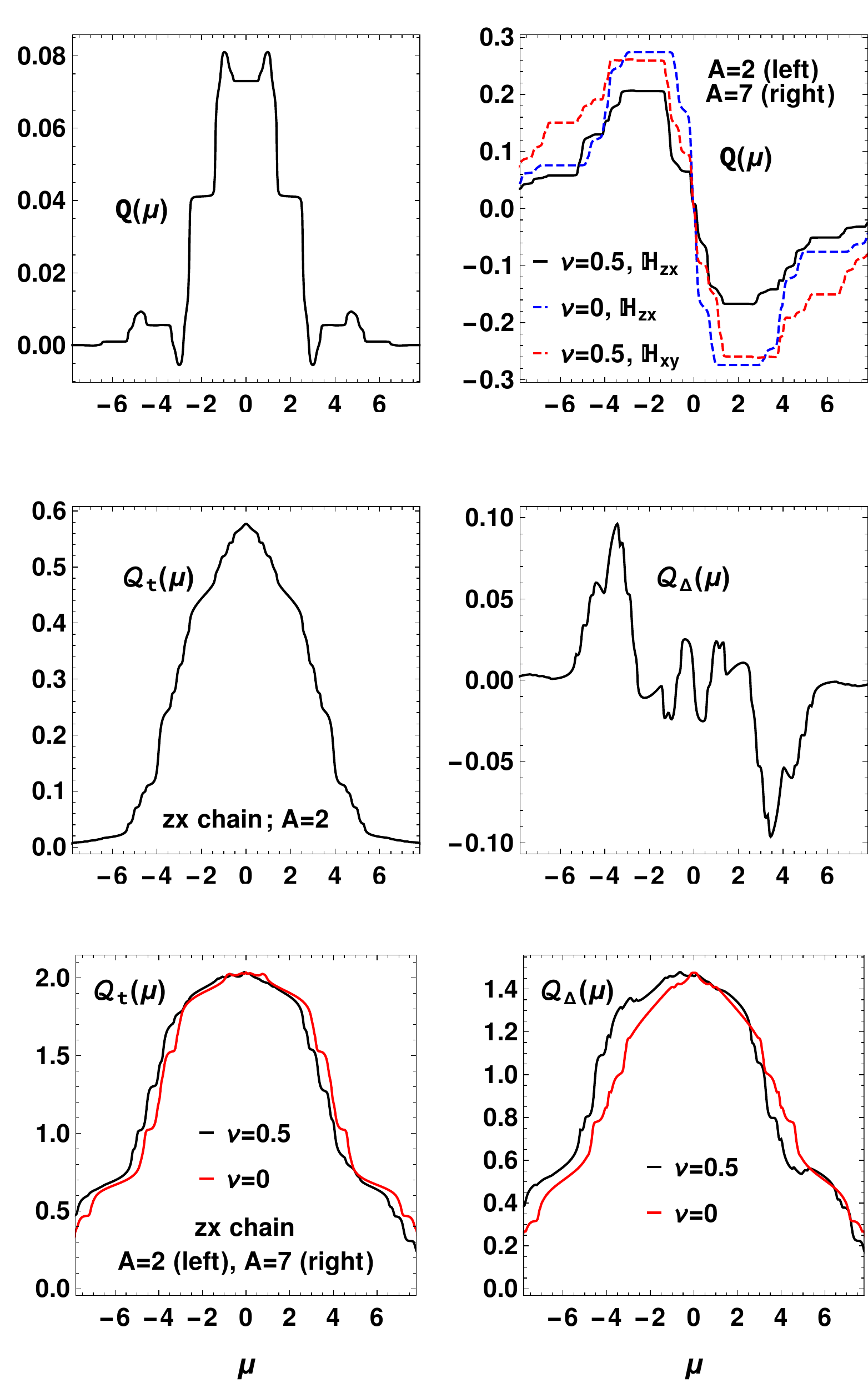}
\end{minipage}
\caption{ Homogeneous (left upper panel) and inhomogeneous (right upper panel) setups, consisting of illuminated SSH chains coupled to wide-band leads. 
Second row: charge pumped in the homogeneous $zx$ chain (left panel) and inhomogeneous (right panel) setups. 
Third row: heat pumping in the homogeneous $zx$ setup.
Bottom row: heat pumping  in the inhomogeneous $zx$ setup.
$\hbar\Omega=2\pi/1.6$, $\gamma=0.5$, $N=10$. 
}
\label{figure1}
\end{figure}

To illustrate the results above, we consider two versions of the SSH model for spinless fermions,
illuminated by monochromatic radiation with angular frequency $\Omega$.  
We study a finite chain with $N/2$ two-atom lattice cells, depicted in Fig.\ref{figure1}-(upper panel), coupled to two infinite wide-band leads.

In the wide-band approximation, the real part of the self-energy vanishes and
the imaginary part becomes energy-independent~\cite{hanggi}. 
This simplification allows to compute the Green's function in a rather explicit way. 
Recalling the definition of the effective Hamiltonian,
we obtain $Z\left(t\right)=\mathbb{H}\left(t\right)-\frac{i}{2}\left(\Gamma_{\t L}+\Gamma_{R}\right)$,
where $\mathbb{H}(t)$ is the time-periodic single-particle Hamiltonian.
The eigenstates of this operator obey the Floquet equation, 
\begin{eqnarray}
i\hbar\partial_{t}|\phi_{\epsilon}(t)\rangle=\left[Z(t)-\epsilon\right]|\phi_{\epsilon}(t)\rangle\,,
\end{eqnarray}
with $\epsilon$ the quasi-energy. The time Fourier series for the
Floquet state reads 
\begin{eqnarray}
|\phi_{\epsilon}(t)\rangle=\sum_{n\in\mathbb{Z}}e^{-in\Omega t}|\Phi_{n}(\epsilon)\rangle\,.
\end{eqnarray}
Expanding the effective Hamiltonian as $Z(t)=\sum_{n}Z_{n}e^{in\omega t}$,
the Fourier components of the Floquet state, $|\Phi_{n}(\epsilon)\rangle$,
satisfy the equation 
\begin{eqnarray}
\sum_{n\in\mathbb{Z}}\left[\ Z_{n-m}-n\hbar\Omega\delta_{n,m}\ \right]|\Phi_{n}(\epsilon)\rangle\ =\epsilon|\Phi_{m}(\epsilon)\rangle\,.
\end{eqnarray}
Because $Z(t)$ is not hermitian, the quasi-energies are, in general,
complex-valued. The Floquet states with $\epsilon$ and $\epsilon+\hbar\omega$
are the physically the same, so it is assumed that $-\hbar\omega/2<\Re(\epsilon)\leq\hbar\omega/2$.
One must also consider the left eigenstates, $\phi_{\epsilon}^{+}(t)$,
satisfying the Floquet equation 
\begin{eqnarray}
-i\hbar\partial_{t}\langle\phi_{\epsilon}^{+}(t)|=\langle\phi_{\epsilon}^{+}(t)|\left[\ Z(t)-\epsilon\ \right]\,,
\end{eqnarray}
whose Fourier time components obey 
\begin{eqnarray}
\sum_{n\in\mathbb{Z}}\langle\Phi_{n}^{+}(\epsilon)|\left[\ Z_{m-n}-m\hbar\Omega\delta_{n,m}\ \right]= &  & \epsilon\langle\Phi_{m}^{+}(\epsilon)|\,,
\end{eqnarray}
and satisfy the normalization condition $\sum_{n}\langle\Phi_{n}^{+}(\epsilon)|\Phi_{n}(\epsilon)\rangle=1$.
The orthonormality and completeness of the right- and left- eigenvector
basis works out for the lattice sites $|i\rangle$ as 
\begin{eqnarray}
\sum_{\epsilon}\sum_{m\in\mathbb{Z}}\ \ket{\Phi_{m}^{+}(\epsilon)}\bra{\Phi_{m}(\epsilon)}=1.
\end{eqnarray}
Using this relations, we obtain the Green's function for Floquet systems as
\begin{eqnarray}
{\cal G}_{\left(m\right)}(E)=\sum_{\epsilon}\sum_{n}\frac{\ket{\Phi_{m+n}^{+}(\epsilon)}\bra{\Phi_{n}(\epsilon)}}{E-\epsilon-n\hbar\Omega}\,.\label{GreenF}
\end{eqnarray}

The two versions of the SSH model we shall consider read, in momentum space, as:
\begin{eqnarray}
\mathbb{H}_{xy}(k,t) & = & \left[\cos k+\nu+A\cos(\Omega t),\sin k,0\right]\cdot\vec{\sigma}\,,\label{xy}\\
\mathbb{H}_{zx}(k,t) & = & \left[\sin k,0,\cos k+\nu+A\cos(\Omega t)\right]\cdot\vec{\sigma}\,,\label{zx}
\end{eqnarray}
where the three Pauli matrices $\vec{\sigma}$ act on sublattice space
and $k$ is the Bloch wave vector over lattice cells.
$\mathbb{H}_{xy}$  and $\mathbb{H}_{zx}$ only differ through a rotation in  $\vec\sigma$ space, 
therefore, they belong in the same symmetry class BDI.
Note that, for an infinite chain, both Hamiltonians obey 
PS as $\mathbb{U}_P\mathbb{H}(-k,t) \mathbb{U}_P^\dagger= \mathbb{H}(k,t)$,
with  $\mathbb{U}_P=\sigma_1$ for $\mathbb{H}_{xy}$,   and 
$\mathbb{U}_P=\sigma_3$ for $\mathbb{H}_{zx}$.

For the chains in Fig.\ref{figure1}, the total self-energy then reads
\begin{eqnarray}
\Sigma&=& -\frac{i}{2}\ \gamma_L|1\rangle\langle 1| -\frac{i}{2} \ \gamma_R  |N\rangle\langle N|
\,,\label{selfenergy}
\end{eqnarray}
where $|j\rangle$  denotes the state at site $j$.  We note that $|1\rangle$ is the first site of the cell
at $x=1$ and  $|N\rangle$ is the second site of the cell  at $x=N/2$.
In the following we shall take $ \gamma_L= \gamma_R=\gamma$.

A finite chain described by $Z_{xy}=\mathbb{H}_{xy}(t)+\Sigma$ still enjoys PS. 
When applying the parity transformation, we note that if the cell $x=(|1\rangle,|2\rangle)$ then the
parity transformation means that we take the cell $-x=(|N-1\rangle,|N\rangle)$. 
Using  $ \mathbb{U}_P=\sigma_1$  we get
\begin{eqnarray}
 \mathbb{U}_P\Sigma(-x)\mathbb{U}_P^\dagger &=& 
  -\frac{i}{2}\gamma \sigma_1\left[  \ |N-1\rangle\langle N-1| + |2\rangle\langle 2| \ 
  \right] \sigma_1\nonumber\\ &=& 
  -\frac{i}{2}\gamma\left(  |N\rangle\langle N| + |1\rangle\langle 1|    \right)\ =\   
  \Sigma(x)\,.
  \label{xyHP2}
 \end{eqnarray}  
$Z_{xy}$ also has PHS with $\mathbb{U}_C=\sigma_3$.
Therefore, charge pumping does not occur in an homogeneous $xy$ chain.

For the finite $zx$ chain, described by  $Z_{zx}=\mathbb{H}_{zx}(t)+\Sigma$, 
 $\Sigma$ breaks both  PS and PHS.
This is because, although $\mathbb{H}_{zx}$ admits $\mathbb{U}_P=\sigma_3$ and $\mathbb{U}_C=\sigma_1$, 
we have
\begin{eqnarray}
\mathbb{U}_P\Sigma(-x)\mathbb{U}_P^\dagger   
 \neq  \Sigma(x)\,, \hspace{0.5cm}
\mathbb{U}_C\Sigma^* \mathbb{U}_C^\dagger 
\neq - \Sigma\,.   \label{zxC}
 \end{eqnarray} 
Nevertheless, the product of PS and PHS holds: 
\begin{eqnarray}
\mathbb{U}_P\mathbb{U}_C\Sigma^*(-x)\mathbb{U}_C^\dagger\mathbb{U}_P^\dagger=-\Sigma(x)   \,.
 \end{eqnarray} 
This ${\cal PC}$ symmetry then ensures that  Eqs. (\ref{Qlodd})-(\ref{Qdeltaeven}) hold. So,  
$Q(\mu)$ and  $\mathcal{Q}_{t}\left(\mu\right)$  are even, while $\mathcal{Q}_{\Delta}\left(\mu\right)$
is odd.

We now consider the inhomogeneous system where two halves of the chain are illuminated with  different amplitudes,
as depicted in Fig.\ref{figure1}-(upper-right panel). 
Here, both parity and the $\cal PC$ symmetry are explicitly broken by the non-uniform illumination of the chain.
In the case of the $xy$ chain, PHS still holds, and $Q(\mu)$ is an odd function while $\mathcal{Q}_{t/\Delta}\left(\mu\right)$ is even. 

The inhomogeneous $zx$ chain is invariant under a $XC$ transformation when $\nu=0$, with $X=\mathbb{U} \Pi$,  where, in real space,  $\mathbb{U}|x\rangle=\sigma_3(-1)^x|x\rangle$.  In momentum space, $\mathbb{U}=\sigma_3\otimes(k\rightarrow k+\pi)$.
Setting $\mathbb{U}_C = 1$, $X C$  transforms the Hamiltonian as
 \begin{eqnarray}
X\left[ \mathbb{H}_{zx}^*(-k,t)\right]_\nu X^{-1}&=& -\left[  \mathbb{H}_{zx}(k,t)\right]_{-\nu}
\label{OHzx}\,,\label{Osigma1}\\
X\Sigma^*X^{-1}&=& -\Sigma\,.
\label{OSigma}  
 \end{eqnarray}
Therefore, PHS, implemented by $X C$, holds for  $\nu=0$ and renders $Q(\mu)$ odd and $\mathcal{Q}_{t/\Delta}\left(\mu\right)$ even. 
Because the homogeneous $zx$ chain, for $\nu=0$, enjoys both the above $\cal PC$ symmetry and the PHS of Eqs. (\ref{OHzx}) and (\ref{OSigma}),  no charge or heat pumping occurs.
We note, for the sake of completeness, that the inhomogeneous $xy$ chain also enjoys a similar PHS  for $\nu=0$, 
but with $\mathbb{U}|x\rangle=(-1)^x|x\rangle$. 

Table \ref{default} summarizes these results for the model systems considered. 
Some representative cases of charge and heat pumping are also illustrated in Figure \ref{figure1},  
exhibiting the even/odd parity identified above.
Note that for the inhomogeneous $zx$ chain with $\nu\neq 0$, neither
the charge (black line in the second row right panel) or heat (bottom panels red line) pumping are odd or even,
as none of the above discussed symmetries exist.

\section{Summary}\label{SecSummary}

In summary, we have discussed the role of discrete symmetries on the pumping of charge, energy or heat.
PHS causes the charge (heat) pumping to be an odd  (even) function 
of the chemical potential. 
On the other hand,  the composition of PS and PHS
causes the charge (heat) pumping to be an even 
(odd) function, and the total heat absorbed to be even.

For the case where the exact symmetry is broken 
the curves do not depart significantly from even/odd functions, as shown  in Fig 1 for the $zx$ chain with $\nu=0.5$.
We then  expect that if the symmetries do not exactly hold (because of interactions, for instance),
the charge/heat pumped currents would still resemble even or odd functions, as we predict,
as long as the system remains  well described by a low energy particle-hole symmetric Hamiltonian.

These results provide simple practical criteria to control the direction of
the charge or heat flows, following the symmetry properties of physical setups.

\section*{Acknowledgments}

We acknowledge partial support from
Funda\c{c}\~ao para a Ci\^encia e Tecnologia (Portugal) through Grant No.
UID/CTM/04540/2019. 

\appendix

\section{Symmetry properties of Floquet scattering matrix}
\label{ApFloS}

We now prove the symmetry properties of the Floquet scattering matrix presented in the main text. 
\begin{enumerate}
\item  TRS:  There exists a unitary matrix, $\mathbb{U}_T$, such that  $\mathbb{U}_T\psi^*(x,-t)$ has the same quasi-energy, $E$.  
Complex conjugation with $t\rightarrow -t$ does not change the time-dependent exponentials, 
but the spatial part is modified  as
 \begin{eqnarray}
\mathbb{U}_T\phi_n^*(x\rightarrow -\infty) &=& A_n^*\mathbb{U}_T e^{-ikx} + C_n^* \mathbb{U}_T e^{ikx}\\
\mathbb{U}_T\phi_n^*(x\rightarrow +\infty) &=& D_n^*\mathbb{U}_T e^{-ikx} + B_n^* \mathbb{U}_T e^{ikx}
\,,
\end{eqnarray}
(here it is assumed that $\mathbb{U}_T$ acts on the spinor form of the plane waves). 
This operation inverts the  direction of propagation of the plane waves.
 We then write  
 \begin{eqnarray}
\left[\begin{array}{c}  A_{n'}^* \\  B_{n'}^*   \end{array} \right]
 &=& S(E+n'\hbar\Omega, E+n\hbar\Omega)\
\left[\begin{array}{c}  C_n^*\\  D_n^*  \end{array} \right]
\,,\label{b7}
\end{eqnarray}
Then, from (\ref{b7}) and (\ref{Sdefin}) we see that
\begin{eqnarray}
\left[\begin{array}{c}  A^* \\  B^*   \end{array} \right] =S\
\left[\begin{array}{c}  C^*\\  D^*  \end{array} \right]
\Leftrightarrow\ S^\top \left[\begin{array}{c}  C^*\\  D^*  \end{array} \right] 
=S \left[\begin{array}{c}  C^*\\  D^*  \end{array} \right] 
\,.\nonumber\\
\end{eqnarray}
Thus, 
$ S^\top=S$.

 \item PHS: There exists a unitary matrix, $\mathbb{U}_C$, such that the state  $\mathbb{U}_C\psi^*(x,t)$ has quasi-energy
  $-E$. 
  Note that complex conjugation changes both the time and spatial dependence of the exponentials, 
  therefore, the direction of propagation of the waves is not changed. The state  $\mathbb{U}_C\psi^*(x,t)$ has
  the asymptotic behavior:
 \begin{eqnarray}
 \begin{split}
 &\sum_n\left( A_n^*\mathbb{U}_C e^{-ikx} + C_n^* \mathbb{U}_C e^{ikx}\right) e^{\frac{i}{\hbar}Et}e^{in\Omega t}\\
 &\hspace{5cm}\mbox{as}\ x\rightarrow -\infty\,,\\ \\
& \sum_n \left( D_n^*\mathbb{U}_C e^{-ikx} + B_n^* \mathbb{U}_C e^{ikx}\right) e^{\frac{i}{\hbar}Et}e^{in\Omega t}\\
 &\hspace{5cm}\mbox{as}\ x\rightarrow \infty\,.
 \end{split}
 \nonumber\\
 \label{778}
\end{eqnarray}
The waves `$n$'' have energy $-E-n\hbar\Omega$. 
Taking the energy labels into account and  
 the definition of the $S$ matrix, we write 
  \begin{eqnarray}
 \left[\begin{array}{c}  C_{n'}^*\\  D_{n'}^*  \end{array} \right] &=& S(-E-n'\hbar\Omega,-E-n\hbar\Omega)
  \left[\begin{array}{c}  A_{n}^*\\  B_{n}^*  \end{array} \right]
 \,,
  \nonumber\\ \label{s14}
\end{eqnarray}
and comparing with  Eq.(\ref{Sdefin}) we get 
\begin{eqnarray}
\begin{split}
& S^*(E+n'\hbar\Omega,E+n\hbar\Omega) =  \\
& S(-E-n'\hbar\Omega,-E-n\hbar\Omega) \,.\\
\end{split}
\end{eqnarray}
In particular, for the case  $n=0$, we obtain Eq.(\ref{PHSmatrix}).

\item Parity: There exists a unitary matrix, $\mathbb{U}_P$, such that
the function $P\psi(-x,t)$ has the same quasi-energy, $E$.  The function $P\psi(-x,t)$ then obeys 
 \begin{eqnarray}
\mathbb{U}_P\phi_n(x\rightarrow -\infty) &=& D_n\mathbb{U}_P e^{ikx} + B_n\mathbb{U}_P e^{-ikx} \\
\mathbb{U}_P\phi_n(x\rightarrow +\infty) &=& A_n\mathbb{U}_P e^{ikx} + C_n\mathbb{U}_P e^{-ikx}
\,,
\end{eqnarray}
so, we write
 \begin{equation}
\left[\begin{array}{c}  D\\  C  \end{array} \right] 
=S \left[\begin{array}{c} B\\  A \end{array} \right]
\Leftrightarrow\ \sigma_1 S\sigma_1 = S \,,
\end{equation}
where $\sigma_1$ acts on the  $(L,R)$ subspace.  The blocks of the $S$ matrix
 in (\ref{Sblocks}) then obey Eqs.(\ref{SLLsymm})-(\ref{SLRsymm}).
\end{enumerate}

\section{Green's function under symmetry transformations}
\label{ApA}

We first consider the total Hamiltonian, i.e. system + leads,
$H\left(t\right)=\frac{1}{2}\bs C^{\dagger}\mathbb{H}\left(t\right)\bs C$,
with $\mathbb{H}^{\dagger}\left(t\right)=\mathbb{H}\left(t\right)$,
and obtain the transformation properties of the total Green's function.
In this case, the evolution of the operator $\bs C$ under the transformed
Hamiltonian, $H_{X}\left(t\right)=\frac{1}{2}\bs C^{\dagger}\mathbb{H}_{X}\left(t\right)\bs C$,
is 
\begin{align}
\pd_{t}\bs C_{X}\left(t\right) & =i\left[H_{X}\left(t\right),\bs C\left(t\right)\right]=-i\mathbb{H}_{X}\left(t\right)\bs C\left(t\right)\,,
\end{align}
and thus
\begin{align}
\bs C_{X}\left(t\right) & =\mathbb{U}_{X}\left(t,0\right)\bs C\,,
\end{align}
with
\begin{align}
\mathbb{U}_{X}\left(t,t'\right) & =\begin{cases}
\mathcal{T}e^{-i\int_{t'}^{t}d\tau\ \mathbb{H}_{X}\left(\tau\right)} & \t{for}\ t>t'\\
\mathbb{U}_{X}\left(t',t\right)^{\dagger} & \t{for}\ t'>t
\end{cases}\,,
\end{align}
yielding the retarded Green's function 
\begin{equation}
G_{X}^{R}\left(t,t'\right)=-i\Theta\left(t-t'\right)\mathbb{U}_{X}\left(t,t'\right).\label{eq:transformed Green's function}
\end{equation}
Under time reversal,
\begin{align}
\mathbb{H}_{T}\left(t\right) & =\mathbb{U}_{T}^{\dagger}\mathbb{H}^{*}\left(-t\right)\mathbb{U}_{T}\,,
\end{align}
and therefore
\begin{align}
G_{T}^{R}\left(t,t'\right)= & -i\Theta\left(t-t'\right)\mathcal{T}e^{-i\int_{t'}^{t}d\tau\ \mathbb{H}_{T}\left(\tau\right)}\nonumber \\
= & \mathbb{U}_{T}^{\dagger}\left[-i\Theta\left(t-t'\right)\mathcal{T}e^{-i\int_{-t}^{-t'}d\tau\ \mathbb{H}\left(\tau\right)}\right]^\top\mathbb{U}_{T}\nonumber \\
= & \mathbb{U}_{T}^{\dagger}\left[G^{R}\left(-t',-t\right)\right]^\top\mathbb{U}_{T},
\end{align}
where $\mathcal{T}$ and $\bar{\mathcal{T}}$ are, respectively, the forward-time ordered and backward-time ordered operators.  

Under charge conjugation,
\begin{align}
\mathbb{H}_{C}\left(t\right)= & -\mathbb{U}_{C}^\top\mathbb{H}^\top\left(t\right)\mathbb{U}_{C}^{*}\,,
\end{align}
we get
\begin{align}
G_{C}^{R}\left(t,t'\right)= & -i\Theta\left(t-t'\right)\mathcal{T}e^{-i\int_{t'}^{t}d\tau\ \mathbb{H}_{C}\left(\tau\right)}\nonumber \\
= & -\mathbb{U}_{C}^\top\left[-i\Theta\left(t-t'\right)\mathcal{T}\prod_{\tau=t'}^{t}e^{-i\mathbb{H}\left(t\right)\Delta t}\right]^{*}\mathbb{U}_{C}^{*}\nonumber \\
= & -\mathbb{U}_{C}^\top\left[G^{R}\left(t,t'\right)\right]^{*}\mathbb{U}_{C}^{*}\,.
\end{align}
Under time translation by half a period we have,
\begin{align}
\mathbb{H}_{\Pi}\left(t\right)= & \mathbb{U}_{\Pi}^{\dagger}\mathbb{H}\left(t+\frac{T}{2}\right)\mathbb{U}_{\Pi}\,,
\end{align}
which yields
\begin{align}
G_{\Pi}^{R}\left(t,t'\right)= & -i\Theta\left(t-t'\right)\mathcal{T}e^{-i\int_{t'}^{t}d\tau\ \mathbb{H}_{\Pi}\left(\tau\right)} \nonumber \\
= & \mathbb{U}_{\Pi}^{\dagger}\left[-i\Theta\left(t-t'\right)\mathcal{T}\prod_{\tau=t'}^{t}e^{i\mathbb{H}\left(t+\frac{T}{2}\right)\Delta t}\right]\mathbb{U}_{\Pi} \nonumber \\
= & \mathbb{U}_{\Pi}^{\dagger}\left[G^{R}\left(t+\frac{T}{2},t'+\frac{T}{2}\right)\right]\mathbb{U}_{\Pi}\,.
\end{align}
Finally, under $x$-coordinate inversion,
\begin{align}
\bra xG_{R}^{R}\left(t,t'\right)\ket{x'}= & \bra{-x}\mathbb{U}_{R}^{\dagger}G_{R}^{R}\left(t,t'\right)\mathbb{U}_{R}\ket{-x'}.
\end{align}

If the unitary matrices, $\mathbb{U}_{X}$, do not mix degrees of freedom of the system with those of the leads, 
we may consider 
Green's functions restricted to degrees of freedom solely within the system by defining
$G_{\t S}^{R}=\sum_{\alpha,\beta\in\t S}\ket{\alpha}\bra{\alpha}G^{R}\ket{\beta}\bra{\beta}$.
However, rather than respecting Eq.(\ref{eq:transformed Green's function}),
the Green's function obeys 
\begin{align}
\left[i\pd_{t}-\mathbb{H}_{X,\t S}\left(t\right)\right]G_{X,\t S}^{R}\left(t,t'\right)  =  \hspace{3.0 cm} \nonumber \\ 
\hspace{2.0 cm} \int d\tau\sum_{l=\t{R,L}}\Sigma_{X,l}^{R}\left(t,\tau\right)G_{X,\t S}^{R}\left(\tau,t'\right),
\end{align}
where $\Sigma_{X,l}^{R}$ is the transformed self-energy obeying
the same $G_{X,\t S}^{R}$. In the following, as in the main text,
we drop the the label $\t S$ and refer to $G^{R}$ as the Green's
function of the system. 

For periodically driven systems, the Floquet retarded Green's function is
defined as 
\begin{align}
& G_{mm'}^{R}\left(\epsilon\right) =  \\
 & \frac{1}{T}\int_{0}^T dte^{i\left(\Omega_{m}-\Omega_{m'}\right)t}\int d\tau G^{R}\left(t,t+\tau\right)e^{-i\left(\epsilon+\Omega_{m'}\right)\tau},  \nonumber 
\end{align}
where $\Omega_{m}=m\Omega$, $m\in\mathbb{Z}$. There is some redundancy
in the definition of this quantity,  since $G_{m+k,m'+k}^{R}\left(\epsilon-\Omega_{m}\right)=G_{m,m'}^{R}\left(\epsilon\right)$,
which is lifted by defining the Floquet Green's function given in
the main text by 
\begin{align}
\mathcal{G}_{\left(m\right)}\left(\epsilon\right)= & G_{m0}^{R}\left(\epsilon\right).
\end{align}
Nevertheless, it is useful to consider $G_{mm'}^{R}\left(\epsilon\right)$
for deriving intermediate expressions.  

Under the time reversal transformation we have 
\begin{align}
&G_{T,mm'}^{R}\left(\epsilon\right)   = \nonumber \\
&=  \frac{1}{T}\int_{0}^\top dte^{i\left(\Omega_{m}-\Omega_{m'}\right)t}\int d\tau G_{T}^{R}\left(t,t+\tau\right)e^{-i\left(\epsilon+\Omega_{m'}\right)\tau}\nonumber \\
 &= \mathbb{U}_{T}^{\dagger}\left[G_{m'm}^{R}\left(\epsilon\right)\right]^\top \mathbb{U}_{T},
\end{align}
using $t-\tau=t'+nT$, $n\in\mathbb{Z}$. Similarly, for charge
conjugation 
\begin{align}
G_{C,mm'}^{R}\left(\epsilon\right) 
 & =-\mathbb{U}_{C}^\top \left[G_{-m-m'}^{R}\left(-\epsilon\right)\right]^{*}\mathbb{U}_{C}^{*},
\end{align}
and for the time translation by half a period, 
\begin{align}
G_{\Pi,mm'}^{R}\left(\epsilon\right) 
 & =\left(-1\right)^{m+m'}\mathbb{U}_{\Pi}^{\dagger}G_{mm'}^{R}\left(\epsilon\right)\mathbb{U}_{\Pi}.
\end{align}
Using the definition of the Floquet Green's function,
we obtain the transformations given in the main text. 

\section{Transmission Probabilities under symmetry transformations \label{ApB} }

\subsection{$T$, $C$ and  $P$}

Using the transformation properties of $\mathcal{G}_{\left(m\right)}\left(\epsilon\right)$
and $\Gamma_{l}\left(\epsilon\right)$, the transmission probabilities
under $T$ transform as
\begin{align}
 & T_{T;ll'}^{\left(m\right)}\left(\epsilon\right)=\nonumber \\
= & \tr\left[\mathcal{G}_{\left(-m\right)}^\top \left(\epsilon+\Omega_{m}\right)\Gamma_{l'}^\top \left(\epsilon\right)\mathcal{G}_{\left(-m\right)}^{\dagger T}\left(\epsilon+\Omega_{m}\right)\mathcal{G}_{l}^\top \left(\epsilon+\Omega_{m}\right)\right]\nonumber \\
= & T_{l'l}^{\left(-m\right)}\left(\epsilon+\Omega_{m}\right)^{*},
\end{align}
and, similarly, under $C$,
\begin{align}
 & T_{C;ll'}^{\left(m\right)}\left(\epsilon\right)=\nonumber \\
= & \tr\left[\mathcal{G}_{C,\left(-m\right)}^{*}\left(-\epsilon\right)\Gamma_{C,l'}^{*}\left(-\epsilon\right)\mathcal{G}_{C\left(-m\right)}^{\dagger*}\left(-\epsilon\right)\Gamma_{C,l}^{*}\left(-\epsilon-\Omega_{m}\right)\right]\nonumber \\
= & T_{ll'}^{\left(-m\right)}\left(-\epsilon\right).
\end{align}
 
 Invariance under $x$-axis inversion implies that the
Green's function is invariant but the hybridization matrices are mapped
onto each other, i.e.,
\begin{align}
\mathcal{G}_{P;\left(m\right)}\left(\epsilon\right)= & \mathcal{G}_{\left(m\right)}\left(\epsilon\right),\\
\Gamma_{P,l}\left(\epsilon\right)= & \Gamma_{\bar{l}}\left(\epsilon\right).
\end{align}
In this case, 
\begin{align}
T_{ll'}^{\left(m\right)}\left(\epsilon\right) & =\tr\left[\mathcal{G}_{\left(m\right)}\left(\epsilon\right)\Gamma_{\bar{l'}}\left(\epsilon\right)\mathcal{G}_{\left(m\right)}^{\dagger}\left(\epsilon\right)\Gamma_{\bar{l}}\left(\epsilon+m\hbar\Omega\right)\right]\nonumber \\
 & =T_{\bar{l}\bar{l'}}^{\left(m\right)}\left(\epsilon\right).
\end{align}

\section{The Role of Time Reversal}\label{ApC}

We here consider the role of time reversal symmetry on the charge
and energy currents. Introducing the transformation of the transmission
probability, given in the main text, into the expression for the particle current,
we obtain 
\begin{align}
J_{l}^{c} & =-\frac{|e|}{h}\sum_{m\in\mathbb{Z}}\int d\epsilon\frac{1}{2}\left\{ T_{l\bar{l}}^{\left(m\right)}\left(\epsilon\right)f_{\bar{l}}\left(\epsilon\right)\right.\nonumber\\
&\hspace{2cm} \left. -T_{l\bar{l}}^{\left(-m\right)}\left(\epsilon+m\hbar\Omega\right)f_{l}\left(\epsilon\right)\right\} \nonumber \\
 & =-\frac{|e|}{h}\sum_{m\in\mathbb{Z}}\int d\epsilon\frac{1}{2}\left\{ T_{l\bar{l}}^{\left(m\right)}\left(\epsilon\right)\left[f_{\bar{l}}\left(\epsilon\right)-f_{l}\left(\epsilon+m\hbar\Omega\right)\right]\right\}\,.
\end{align}
Similarly, for the energy current, 
\begin{align}
J_{l}^{e} & =\frac{1}{h}\sum_{m\in\mathbb{Z}}\int d\epsilon\left\{ \left(m\hbar\Omega+\epsilon\right)T_{l\bar{l}}^{\left(m\right)}\left(\epsilon\right)\left[f_{\bar{l}}\left(\epsilon\right)-f_{l}\left(m\hbar\Omega+\epsilon\right)\right]\right.\nonumber \\
 & \left.+\frac{1}{2}m\hbar\Omega T_{ll}^{\left(m\right)}\left(\epsilon\right)\left[f_{l}\left(\epsilon\right)-f_{l}\left(\epsilon+m\hbar\Omega\right)\right]\right\} .
\end{align}
Therefore, for the pumping setup $f_{l}\left(\epsilon\right)=f\left(\epsilon\right)$,
we find that only in the infinite temperature case can the time-reversed
processes happen with the same probability. In that case $f_{l}\left(\epsilon\right)=\frac{1}{2}$
and $J_{l}^{c}=J_{l}^{e}=0.$

\bibliographystyle{apsrev4-1}
\bibliography{references}

\end{document}